\documentclass[12pt]{iopart}
\usepackage{iopams}
\usepackage{amsfonts}
\usepackage{graphicx}
\usepackage{epsfig}
\usepackage{graphics}
\makeatletter
\newcommand{\row}[1]%
{\mathord{\buildrel{\lower3pt%
\hbox{$\scriptscriptstyle\rightarrow$}}\over #1}}

\newcommand{\dyadic}[1]{\mathord{\dyadic@rrow{#1}}}
\newcommand{\dyadic@rrow}[1]{
\begin{picture}(12,12)(-1,0)
\put(-2,10){\makebox(0,0)[t]{$\scriptscriptstyle\downarrow$}}
\put(-2,11){\makebox(0,0)[l]{$\scriptscriptstyle\longrightarrow$}}
\put(5,0){\makebox(0,0)[b]{$#1$}}
\end{picture}
}

\newcommand{\bra}[1]{\bigl\langle #1 \bigr|}
\newcommand{\ket}[1]{\bigl| #1 \bigr\rangle}

\makeatletter

\begin{document}

\title{ Quantum coding  in non-inertial frames }
\author{ N. Metwally$^{(1,2)}$ and A. Sagheer$^{(1)}$\\}
\address
{ $^{1,2}$Math. Dept., Faculty of Science, Aswan university,
Aswan,
Egypt.\\
$^{2}$Mathematics  Department, College of Science, University of Bahrain,\\
P. O. Box 32038 Kingdom of Bahrain.}
 \ead{nasser@Cairo-aswu.edu.eg}

 \begin{abstract}
The capacity of accelerated  channel is investigated for different
classes of initial states. It is shown that, the capacities of the
travelling channels depend on  the frame in which the accelerated
channels are observed in  and the initial shared state between the
partners. In some frames, the capacities decay as the
accelerations of both qubit increase. The decay rate is larger if
the partners are initially share a maximum entangled state. The
possibility of using the accelerated quantum channels to perform
quantum coding protocol is discussed. The  amount of decoded
information  is quantified for different cases, where it decays as
the partner's accelerations increase to reach its minimum bound.
This minimum bound depends on the initial shared states and it is
large for maximum entangled state.
 \end{abstract}
\maketitle

\section{Introduction}
Nowadays, investigating the dynamics of quantum states in non
inertial frames  represents one of the most events in the context
of quantum information. For example, M. del Rey et al.\cite{Marco}
have presented a scheme for simulating a set of accelerated atoms
coupled to a single mode field. The dynamics of multipartite
entanglement of fermionic systems in non-inertial frames is
discussed by Wang and Jing \cite{Wang}. The effect of deoherence
on a qubit-qutrit system is studied by Ramzan and Khan
\cite{Ramazan}. The influence of Unruh effect on the payoff
function of the players  in the quantum Prisoners is investigated
in \cite{khan}. Goudarzi and Beyrami \cite{Goud} have discussed
the effect of uniform acceleration on multiplayer quantum game.
The usefulness  classes of travelling entangled quantum channels
in non-inertial frames is investigated by Metwally
\cite{metwally1}.

Manipulating some quantum information tasks in these non-inertial
frames represents a real desired aim. Therefore some efforts have
been done to investigate quantum teleportation via accelerated
states as quantum channels. For example, Alsing and  Milburn
\cite{Alsing}gave a description  of the quantum teleportation
between two users, one of them in an inertial frame and the other
moves in an accelerated frame.
 In \cite{Metwally2}, the possibility
of using maximum and partial entangled states as an initial states
to perform quantum teleportation in non-inertial frames is
discussed, where it is assumed that  the teleported information
can be either  accelerated or non accelerated.

Quantum coding protocols  represent one of the most important ways
to send coded information between two users\cite{Ben1,Bos,Bown}.
Therefore, it is important to discuss the possibility of using
different accelerated  states, which are initially prepared in
maximum or partial entangled states to perform quantum coding
protocol. In this contribution, we investigate the behavior of the
capacities of different accelerated states. Theses states  have
been used as quantum channels to implement the original quantum
coding protocol\cite{Ben1}.

The paper is organized as follows: In Sec.2,  the initial system
is described, where we assume that the users share a two- qubit
state of X-type \cite{Eberly}. The relation between Minkowski and
Rindler spaces is reviewed. The Unruh effect  on the initial
system is investigated. The channels' capacities are quantified
for different classes of initial states in Sec.3. The accelerated
states are used  as quantum channels to perform quantum coding.
The amount of decoded information  is calculated and the effect of
the accelerations and the initial states setting is discussed in
Sec.4.  Finally, Sec.5 is devoted to discuss our results.

\section{The Model}
 The class of $X-$state \cite{Eberly} represents one of most popular classes of
 quantum channel which have been extensively studied in the
 context of quantum information.  Some properties of this class
have been  discussed in different directions. For example the
phenomena of quantum discord of two-qubit X-states is discussed by
Q. Chen et. al, \cite{Chen,Ali}. The phenomena of entanglement
sudden death of  two-qubit X-states in thermal reservoirs is
discussed in \cite{Rau}. Therefore we are motivated   to
investigate the  behavior of this state in non-inertial
frame\cite{metwally1}. In computational basis
$\ket{00},\ket{01},\ket{10}$ and $\ket{11}$, $X$-state takes the
form,
 \begin{eqnarray}\label{xstate}
 \rho_X&=&\mathcal{A}_{11}\ket{00}\bra{00}+\mathcal{A}_{22}\ket{01}\bra{01}+\mathcal{A}_{33}\ket{10}\bra{10}+
 \mathcal{A}_{44}\ket{11}\bra{11}+\mathcal{A}_{23}\ket{01}\bra{10}
\nonumber\\
 &&+\mathcal{A}_{32}\ket{10}\bra{01}+
 \mathcal{A}_{14}\ket{00}\bra{11}+\mathcal{A}_{41}\ket{11}\bra{00},
 \end{eqnarray}
where
\begin{eqnarray}
\mathcal{A}_{11}&=&\mathcal{A}_{44}=\frac{1}{4}(1+c_z),\quad\mathcal{A}_{22}=\mathcal{A}_{33}=\frac{1}{4}(1-c_z),
\nonumber\\
\mathcal{A}_{23}&=&\mathcal{A}_{32}=\frac{1}{4}(c_x+c_y),\quad\mathcal{A}_{14}=\mathcal{A}_{41}=\frac{1}{4}(c_x-c_y).
\end{eqnarray}
Since,  we are interested to use this class of states to perform
quantum coding in non-inertial frames, it is important to shed
some light on the Minkowski and Rindler spaces in the context of
Dirac field.\\

 \subsection{ Minkowski and Rindler's spaces}
  In the inertial frames, Minkowsik
coordinates $(t,z)$ are  used to  describe Dirac field, while in
the uniformly accelerated case, Rindler coordinates $(\tau, \chi)$
are more adequately. The relations between the Minkowski and
Rindler coordinates  are given by\cite{Alsing,Edu},
\begin{equation}\label{trans}
\tau=r~tanh\left(\frac{t}{z}\right), \quad \chi=\sqrt{t^2-z^2},
\end{equation}
where  $-\infty<\tau<\infty$, $-\infty<\chi<\infty$  and $r$ is
the acceleration of the moving particle. The transformations
(\ref{trans}) define two regions in Rindler's spaces: the first
region $I $ for $|t|<x$  and the second region $II$ for $x<-|t|$ .
A single mode $k$ of fermions and anti-fermions in Minkowski space
is described by the annihilation operators $a_k$  and $b_{-k}$
respectively, where $a_k\ket{0_k}=0$ and
$b^\dagger_{-k}\ket{0_{k}}=0$. In terms of Rindler's operator(
$c^{(I)}_k, d^{(II)}_{-k}$), the Minkowski operators can be
written as \cite{Walls},
\begin{equation}\label{op}
a_k=\cos r c^{(I)}_k-\exp(-i\phi)\sin r d^{(II)}_{-k}, \quad
b^\dagger_{-k}=\exp(i\phi)\sin r c^{(I)}_k+\cos r d^{(II)}_{k},
\end{equation}
where $tan r=e^{-\pi\omega \frac{c}{a}}$, $0\leq r\leq \pi/$4, $
a$ is the acceleration such that $0\leq a\leq\infty$, $\omega$ is
the frequency of the travelling qubits, $c$ is the speed of light
and $\phi$ is an unimportant phase that can be absorbed into the
definition of the operators. It is clear that, the transformation
(\ref{op}) mixes a particle in region $I$ and an anti particle in
region $II$. This effect is called Unruh effect \cite{Alsing1,un}.
In terms of Rindler's modes, the  Minkowski vacuum $\ket{0_k}_M$
and the one particle
 states $\ket{1_k}_M$
 take the form,
\begin{eqnarray}\label{Min}
\ket{0_k}_M&=&\cos r\ket{0_k}_I\ket{0_{-k}}_{II}+ \sin
r\ket{1_k}_I\ket{1_{-k}}_{II}, \nonumber\\
\ket{1_k}_M&=&a^\dagger_k\ket{0_k}_M =\ket{1_k}_I\ket{0_k}_{II}.
\end{eqnarray}

\subsection{Unruh effect on X-state}
Since the expressions of the vacuum and single particle states are
obtained in Rindler basis(\ref{Min}), then we can investigate the
dynamics of the suggested state (\ref{xstate}) from the uniformly
accelerated point of view. In this context, using
Eq.(\ref{Min}$\&$\ref{xstate}), one can obtain the form of  the
$X-$ sate in the first  and the second regions.

\begin{eqnarray}
\rho_{x{A_{\ell}B_{\ell}}}&=&\mathcal{B}^{\ell}_{11}\ket{00}\bra{00}+\mathcal{B}^{\ell}_{22}\ket{01}\bra{01}+
\mathcal{B}^{\ell}_{33}\ket{10}\bra{10}+
 \mathcal{B}^{\ell}_{44}\ket{11}\bra{11}+\mathcal{B}^{\ell}_{23}\ket{01}\bra{10}
\nonumber\\
 &&+\mathcal{B}^{\ell}_{32}\ket{10}\bra{01}+
 \mathcal{B}^{\ell}_{14}\ket{00}\bra{11}+\mathcal{B}^{\ell}_{41}\ket{11}\bra{00},
\end{eqnarray}
where, $\ell=I, II $, for the first  and second regions
respectively. If the states of Alice and Bob in the second region
$II$ are traced out, then the accelerated channel in the first
region is defined by the coefficients,
\begin{eqnarray}
\mathcal{B}^{(I)}_{11}&=&\mathcal{A}_{11}\cos^2r_a\cos^2r_b, \quad
\mathcal{B}^{(I)}_{14}=\mathcal{A}_{14}\cos r_a\cos r_b
\nonumber\\
\mathcal{B}^{(I)}_{22}&=&\cos^2r_a(\mathcal{A}_{11}\sin^2r_b+\mathcal{A}_{22}),\quad
\mathcal{B}^{(I)}_{23}=\mathcal{A}_{23}\cos r_a\cos r_b
\nonumber\\
\mathcal{B}^{(I)}_{23}&=&\mathcal{A}_{32}\cos r_a\cos r_b\quad
\mathcal{B}^{(I)}_{33}=\cos^2r_b(\mathcal{A}_{11}\sin^2r_a+\mathcal{A}_{33})
\nonumber\\
\mathcal{B}^{(I)}_{41}&=&\mathcal{A}_{41}\cos r_a\cos r_b, \quad
\mathcal{B}^{(I)}_{44}=\sin^2 r_a(\mathcal{A}_{11}\sin^2
r_b+\mathcal{A}_{22})+\mathcal{A}_{33}\sin^2r_b+\mathcal{A}_{44}
\end{eqnarray}
On the other hand, if we trace out the states of Alice and Bob in
the first region $I$, the accelerated channel  between Alice and
Bob in the second region $II$  is defined by the coefficients,
\begin{eqnarray}
\mathcal{B}^{(II)}_{11}&=&(\mathcal{A}_{22}+\mathcal{A}_{11}\cos^2r_b)\cos^2r_a+\mathcal{A}_{33}\cos^2r_b+\mathcal{A}_{44}
,\quad \mathcal{B}^{(II)}_{14}=\mathcal{A}_{41}\sin r_a\sin r_b
\nonumber\\
\mathcal{B}^{(II)}_{22}&=&(\mathcal{A}_{33+}\cos^2r_a)\sin^2r_b,\quad
\mathcal{B}^{(II)}_{23}=\mathcal{A}_{32}\sin r_a\sin r_b,
\nonumber\\
\mathcal{B}^{(II)}_{23}&=&\mathcal{A}_{23}\sin r_a\sin r_b\quad
\mathcal{B}^{(II)}_{33}=(\mathcal{A}_{22}+\mathcal{A}_{11}\cos^2r_b)\sin^2r_a,
\nonumber\\
\mathcal{B}^{(II)}_{41}&=&\mathcal{A}_{14}\sin r_a\sin r_b, \quad
\mathcal{B}^{(II)}_{44}=\mathcal{A}_{11}\sin^2 r_a\sin^2r_b.
\end{eqnarray}
There are also two channels  that could be investigated,
$\rho_{A_IB_{II}}$ and  $\rho_{A_{II}B_{I}}$ which represent the
channel between Alice, Anti-Bob and Anti-Alice, Bob respectively.
In this context, we consider only the channel between Alice in the
first region $I$ and Bob in the second region $II$.  In this case
the coefficient $\mathcal{B}_{ij}$ are given by,
\begin{eqnarray}\label{aIbII}
\mathcal{B}_{11}&=&(\mathcal{A}_{22}+\mathcal{A}_{11}\cos^2r_b)\cos^2r_a
,\quad \mathcal{B}_{14}=\mathcal{A}_{23}\cos r_a\sin r_b,
\nonumber\\
\mathcal{B}_{22}&=&\mathcal{A}_{11}\cos^2r_a\sin^2r_b,\quad
\mathcal{B}_{23}=\mathcal{A}_{14}\cos r_a\sin r_b,
\nonumber\\
\mathcal{B}_{23}&=&\mathcal{A}_{41}\cos r_a\sin r_b\quad
\mathcal{B}_{33}=(\mathcal{A}_{22}+\mathcal{A}_{11}\cos^2r_b)\sin^2r_a+(\mathcal{A}_{33}\cos^2r_b+\mathcal{A}_{44}),
\nonumber\\
\mathcal{B}_{41}&=&\mathcal{A}_{32}\cos r_a\sin r_b, \quad
\mathcal{B}_{44}=(\mathcal{A}_{33}+\mathcal{A}_{11}\sin^2r_a)\sin^2r_b.
\end{eqnarray}
Since the quantum channels are obtained in the different regions,
it is possible to investigate the behavior of  the  channel
capacity as well as the possibility of using these states as
quantum channels to perform quantum coding.

\section{Channel capacity}

Channel capacity which measures the rate of information transfer
represents one of the most important indictors of the channel's
efficiency.  Therefor, it is necessary to investigate this
phenomena for the travelling channels in different regions.
Mathematically, the channel capacity of the channel $\rho_{A_\ell
B_\ell}, \ell=I,II$ is given by \cite{sohr},
\begin{eqnarray}\label{Cap}
 \mathcal{C_P}&=&\log_aD +(\mathcal{B}^{(\ell)}_{11}+\mathcal{B}^{(\ell)}_{33})\
 log\bigl(\mathcal{B}^{(\ell)}_{11}+\mathcal{B}^{(\ell)}_{33}\bigr)+ \mathcal{B}^{(\ell)}_{22}+\mathcal{B}^{(\ell)}_{44}\
 log\bigl(\mathcal{B}^{(\ell)}_{22}+\mathcal{B}^{(\ell)}_{44}\bigr)
 \nonumber\\
 &&-\lambda_1\log\lambda_1-\lambda_2\log\lambda_2-\lambda_3\log\lambda_1-\lambda_3\log\lambda_3
\end{eqnarray}
where,
 \begin{eqnarray}\label{lambda}
\lambda_{1,2}=\frac{1}{2}\Bigl\{\mathcal{B}^{(\ell)}_{11}+\mathcal{B}^{(\ell)}_{44}\bigr\}
\pm\sqrt{\bigl(\mathcal{B}^{(\ell)}_{11}-\mathcal{B}^{(\ell)}_{44}\bigr)^2+
4\mathcal{B}^{(\ell)}_{41}\mathcal{B}^{(\ell)}_{14}}\Bigr\},
\nonumber\\
\lambda_{3,4}=\frac{1}{2}\Bigl\{\mathcal{B}^{(\ell)}_{22}+\mathcal{B}^{(\ell)}_{33}\bigr\}
\pm\sqrt{\bigl(\mathcal{B}^{(\ell)}_{22}-\mathcal{B}^{(\ell)}_{33}\bigr)^2+
4\mathcal{B}^{(\ell)}_{23}\mathcal{B}^{(\ell)}_{32}}\Bigr\},
\end{eqnarray}
where $D=2$. The channel capacity of the quantum channel between
Alice in the first region and Bob in the second region is given
by(\ref{Cap}), but $\mathcal{B}^{(\ell)}_{ij}$ are given by
(\ref{aIbII}).

\begin{figure}
  \begin{center}
  \includegraphics[width=19pc,height=14pc]{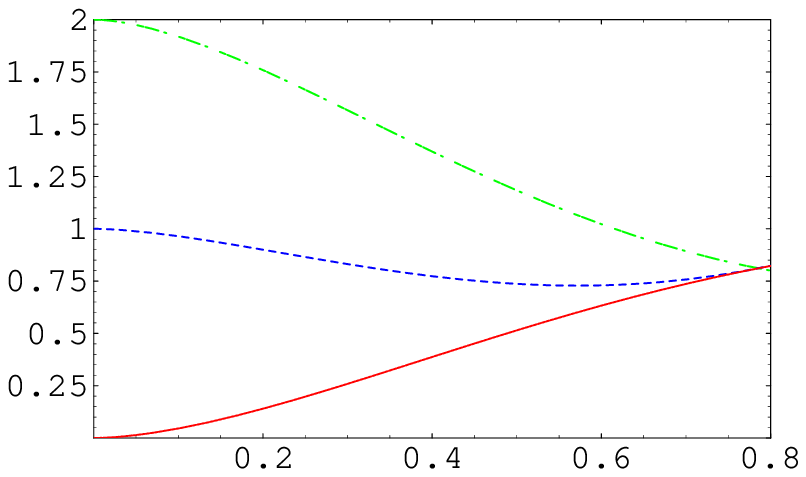}~
     \includegraphics[width=19pc,height=14pc]{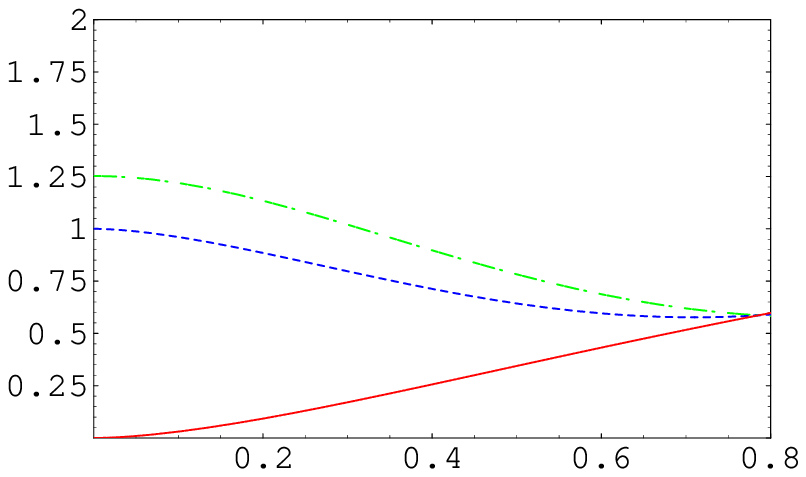}
     \put(-263,150){$(a)$}
 \put(-28,150){$(b)$}
  \put(-100,-5){$r$}
  \put(-340,-5){$r$}
\put(-230,90){$\mathcal{C_P}$}
  \put(-470,90){$\mathcal{C_P}$}
    \caption{The channel's capacity of a system is initially prepared in (a) maximum entangled
state, MES with $c_x=c_y=c_z=-1$  and (b) partial entangled state.
PES with $c_x=-0.9, c_y=-0.8$ and $c_z=-0.7$. The dash dot, dot
and solid curves represent the capacities    of the channels
$\rho_{A_IB_I}, \rho_{A_{II}B_{II}}$ and $\rho_{A_IB_{II}}$
respectively and $r_a=r_b=r$.}
  \end{center}
\end{figure}
The dynamics of the channel capacity of the travelling state is
displayed in Fig.(1) for  a system  initially prepared  in maximum
entangled state.  Fig.(1a), describes the behavior of
$\mathcal{C_P}$ for  quantum channels which are generated between
Alice and Bob in the first region, $\rho_{A_IB_{I}}$, in the
second region $\rho_{A_{II}B_{II}}$ and between Alice in the first
region and Bob in the second region $\rho_{A_IB_{II}}$. In this
investigation, it is assumed that both qubits are accelerated. It
is clear that the  capacity the channel capacity $\mathcal{C_P}$
is maximum  for the channel $\rho_{A_IB_{I}}$ at zero
accelerations. However  as the accelerations of the travelling
qubits increase, the channel capacity decreases smoothly to reach
its minimum values with-$\infty$ acceleration.  In the second
region, the degree of entanglement of the state
$\rho_{A_{II}B_{II}}$ is not maximum \cite{metwally3}, therefore
$\mathcal{C_P}$ is not maximum at zero accelerations. For larger
values of the accelerations the channel's capacity  slightly
decreases. This show that this generated entangled state is more
robust than that in the first region, $\rho_{A_IB_{I}}$. The
generated channel between Alice and Anti-Bob, $\rho_{A_IB_{II}}$
has a zero capacity at $r_a=r_b=0$, since  there is no
entanglement generated between these qubits. However as the
accelerations increase, the channel's capacity $\mathcal{C_P}$
increases gradually to reach its maximum value at -$\infty$
accelerations.

In Fig.(1b), the channel's capacities  of the travelling quantum
channels are investigated  for a system  initially prepared in a
partial entangled state, where we set $c_x=-0.9, c_y=-0.8$ and
$c_z=-0.7$.  It is clear that the general behavior is similar to
that predicted in Fig.(1a), for maximum entangled state, MES.
However in the first region, the initial capacities  of the state
$\rho_{A_IB_I}$ is smaller than that described in Fig.(1a) for
MES. As the accelerations of Alice and Bob's qubit increase the
channel capacity $\mathcal{C_P}$ decreases.  In the second region
$II$, the capacity of the quantum channel which is generated
between Anti-Alice and Anti-Bob, $\rho_{A_{II}B_{II}}$ decreases
smoothly as $r_{a(b)}$ increases. The capacity for the channel
$\rho_{A_IB_{II}}$ increases as the accelerations of both  qubit
increase, but its maximum value is smaller than its corresponding
one  for MES.

From Fig.(1), one concludes that the channel capacity depends on
the initial degree of entanglement of the travelling qubit. In our
recent work \cite{metwally1}, we have shown that the generated
entangled channel in the first region has larger degree of
entanglement than that in the second region. This explains, why
the channel capacity in the first region is much larger than that
displayed in the second region. On the other hand, initially, the
channel between a qubit and the Anti-qubit of the other qubit is
separable.  However  this  channel turns into entangled state as
the accelerations increase and consequently its capacity
increases.

\section{quantum Coding}

In this section, we investigate the possibility of using the
travelling channels between the different users to perform the
quantum coding protocol which is proposed by Bennett and Wienser
\cite{Ben1}.  This protocol works as follows:
\begin{enumerate}
\item Alice encodes the given information in her qubit by using
one of local unitary operators $\mathcal{U}_i=I,\sigma_x,\sigma_y$
or $\sigma_z$ with probability $p_i$. Due to this operation the
information is coded in the state
\begin{equation}
\rho_{cod}=\sum_{i}^{3}\Bigl\{p_i\mathcal{U}_i\otimes
I_2{\rho_x}_{A_{\ell}B_{\ell}}I_2\otimes\mathcal{U}_i^\dagger\Bigl\},
\end{equation}
where $I_2$ is the unitary operator for Bob qubit.

\item Alice sends her qubit to Bob, who makes joint measurements
on the two qubits. The maximum amount of information which Bob can
extract from Alice's message is Bounded by
\cite{metwally3,metwally4},

\begin{equation}
I_{B}=\mathcal{S}(\rho_{cod})-\sum_{i=1}^{3}p_i\mathcal{S}\Bigl(\mathcal{U}_i\otimes
I_2{\rho_x}_{A_{\ell}B_{\ell}}I_2\otimes\mathcal{U}_i^\dagger\Bigr).
\end{equation}
Explicitly,
\begin{equation}
I_B=-2(\lambda_+\log\lambda_{+}+\lambda_{-}\log\lambda_{-})-2(\lambda_{1}\log\lambda_1+
\lambda_{2}
\log\lambda_2+\lambda_3\log\lambda_3+\lambda_4\log\lambda_4),
\end{equation}
where
\begin{eqnarray}
\lambda_{\pm}&=&\frac{1}{2}\Bigl\{1+
\frac{1}{2}\sqrt{\{(\mathcal{B}_{11}^{(\ell)}+\mathcal{B}_{33}^{(\ell)})-
(\mathcal{B}_{22}^{(\ell)}+\mathcal{B}_{44}^{(\ell)})\}^2+
4(\mathcal{B}_{12}^{(\ell)}+\mathcal{B}_{34}^{(\ell)})(\mathcal{B}_{21}^{(\ell)}+\mathcal{B}_{43}^{(\ell)})}~\Bigr\},
\nonumber\\
\end{eqnarray}
and $\lambda_i, i=1..4$ are given in  (\ref{lambda}) and it is
assumed  that Alice has used the unitary operator with an equal
probability i.e., $p_i=\frac{1}{4}$,
$i=I,\sigma_x,\sigma_y,\sigma_z$.

\begin{figure}\label{Fig-BobI}
  \begin{center}
     \includegraphics[width=19pc,height=14pc]{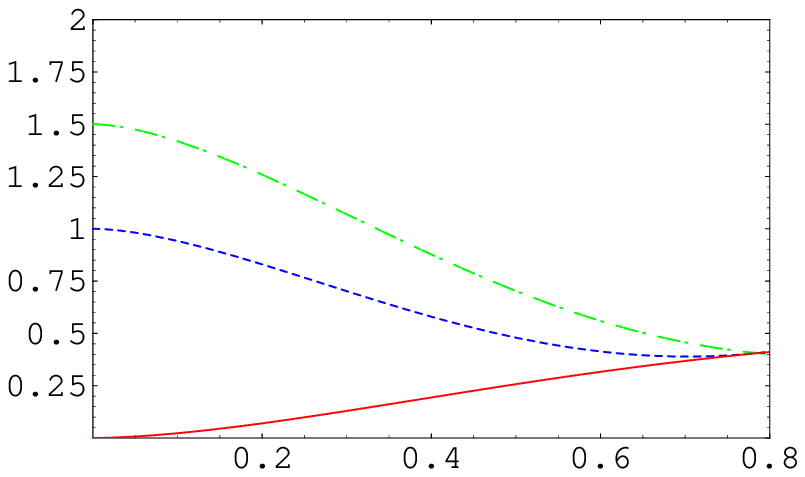}~
      \includegraphics[width=19pc,height=14pc]{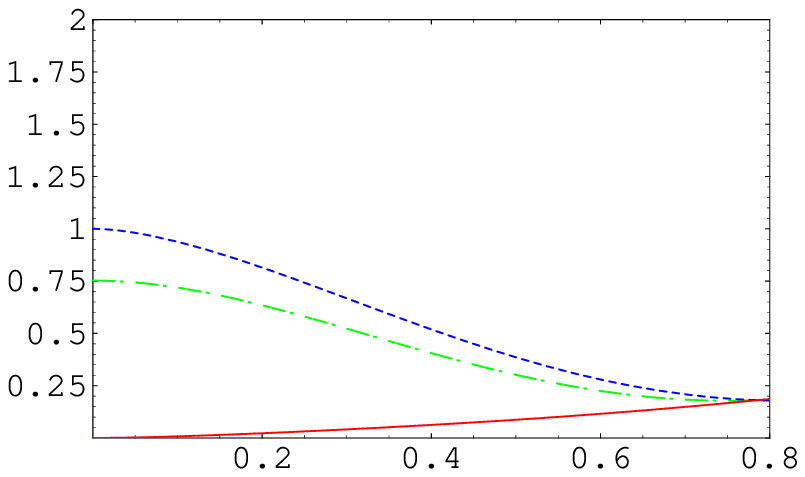}
      \put(-263,150){$(a)$}
 \put(-28,150){$(b)$}
       \put(-470,90){$I_{Bob}$}
\put(-100,-5){$r$}
     \put(-235,90){$I_{Bob}$}
 \put(-340,-5){$r$}
      \caption{The  amount of information decoded by Bob $I_{Bob}$ for (a) MES and (b)PES, where we set the same parameters
as described in Fig.(1).}
  \end{center}
\end{figure}

In Fig.(2), the amount of decoded information by Bob, $I_{Bob}$ is
plotted for different initial  channels. Fig.(2a), describes the
behavior of $I_{Bob}$ for a  qubit system  initially prepared in
maximum entangled state. It is clear that the amount of
information which decoded from the channel between Alice and Bob
in the first region, $\rho_{A_IB_I}$ decreases smoothly to reach
its minimum value when the accelerations tend to infinity. In the
second region the information is coded in the state
$\rho_{A_{II}B_{II}}$. In this case the amount of decoded
information is smaller than that shown in the first region.
However the decoded information from the channel between Alice and
Anti-Bob, $\rho_{A_{I}B_{II}}$ increases as the accelerations
increase  to reach its maximum bound. This maximum bound
represents the lower bound for the decoded information in the
regions $I$ and $II$.

The behavior of the decoded information  in a system  initially
prepared in a partial entangled state is depicted in Fig.(2b). The
general behavior is similar to that shown in Fig.(2a), but the
information decoded by Bob, $I_{Bob}$ is much smaller than  coded
in maximum entangled states. The minimum  bound is smaller than
that displayed  for a system  initially prepared in MES. The
amount of information which is decoded from the quantum  channel
$\rho_{A_{I}B_{II}}$  increases,  with a rate  smaller than its
corresponding one  in  Fig.(2a),as the accelerations increase.

\end{enumerate}

\section{Conclusion}
In this contribution, we investigate the dynamics of the
accelerated channels  in non-inertial frame. It is assumed that
the partners initially share maximum or partial entangled
channels.  The  capacity of the accelerated channels depends on
the initial state setting and the  frame in which the partners are
observed. It is shown that the capacity decays if  both partners
are observed in the same frame, and increases smoothly if the
partners are observed in different frames. The capacity decays
quickly in one region, where the accelerated channel has larger
degree of entanglement. However in some regions, the decay is
small because this accelerated  channel is partially entangled.
This shows that the capacity of larger degree of entanglement
decays faster than that has smaller degree of entanglement.
Starting from a maximum entangled states, the capacity of the
accelerated channels is much larger than that depicted for partial
entangled channel as initial states.

The possibility of  employing the accelerated  channels in
different frames to perform quantum coding is investigated. It is
assumed that the source supplies the partners with  different
classes of initial states. The amount of decoded information is
quantified for different accelerated channels. It is shown the
decoded information decays  as the the accelerations of both qubit
increases. The decay rate  depends on the frame in which the
channel is accelerated, where  in the first region, the rate of
decay is larger than that depicted in the second region.  If the
partners start with a maximum entangled  states, then the decoded
information from the accelerated channels is much larger than that
decoded from accelerated channel generated  initially from partial
entangled channels. The  decoded information from  a channel
between a partner and the Anti-partner increases gradually  to
reach its maximum bound with-$\infty$ acceleration. The increasing
rate depends on the initially shared state between the partners.

{\it In conclusion},  it is possible to use the accelerated
quantum  channels to  send coded information between two partners.
The channel capacity and the amount of decoded information depend
on the frames in which the partners are observed and the  initial
shared state between the partners.

\end{document}